\documentclass[12pt]{article}
\usepackage{amsmath,amssymb}
\usepackage{graphicx}
\usepackage{color}
\usepackage{cite,fancyhdr}
\pdfoutput=1

\usepackage[affil-it]{authblk}
\usepackage[left=2cm,right=2cm,top=3cm,bottom=3cm,a4paper]{geometry}

\begin{document}

\allowdisplaybreaks
\setcounter{footnote}{0}
\setcounter{figure}{0}
\setcounter{table}{0}

\title{\bf \large 
Gauge Coupling Unification with Hidden Photon, \\ and Minicharged Dark Matter}
\author[1]{{\normalsize Ryuji Daido}}
\author[1,2]{{\normalsize Fuminobu Takahashi}}
\author[1]{{\normalsize Norimi Yokozaki}}

\affil[1]{\small 
Department of Physics, Tohoku University,  

Sendai, Miyagi 980-8578, Japan}

\affil[2]{\small 
Kavli Institute for the Physics and Mathematics of the Universe (WPI),
 
University of Tokyo, Kashiwa 277--8583, Japan}

\date{}

\maketitle

\thispagestyle{fancy}
\rhead{TU-1033 \\ IPMU16-0144 }
\cfoot{\thepage}
\renewcommand{\headrulewidth}{0pt}

\begin{abstract}
\noindent
We show that gauge coupling unification is realized with a greater accuracy
 in the presence of a massless hidden photon which has a large kinetic mixing with hypercharge. 
 We solve the renormalization group equations  at two-loop level and find that
the GUT unification scale is around $10^{16.5}$\,GeV which sufficiently 
suppresses the proton decay rate, and that the unification is essentially determined by the kinetic mixing only,
and it is rather insensitive to the hidden gauge coupling or the presence of vector-like matter fields charged under U(1)$_H$ and/or SU(5). 
Matter fields charged under the unbroken hidden U(1)$_H$ are stable and they contribute to dark matter.
Interestingly, they become minicharged dark matter which carries a small but non-zero 
electric charge, if the hidden gauge coupling is tiny. The minicharged dark matter
is a natural outcome of the gauge coupling unification with a hidden photon.
\end{abstract}

\clearpage

\section{Introduction}
The Standard Model (SM) has been so successful that it explains almost all the existing experimental 
data with a very high accuracy. 
The lack of clear evidence for new particles at the LHC experiment
so far began to cast doubt on the naturalness argument which has been the driving force of
search for new physics at TeV scale. 
On the other hand, there are many phenomena that
require physics beyond the SM, such as dark matter,  baryon asymmetry, inflation, 
neutrino masses and mixings, etc.  Among them, the gauge coupling unification in
a grand unified theory (GUT) is an intriguing and  plausible possibility, which has been extensively
studied in the literature.

The running of gauge couplings are obtained by solving the renormalization group (RG) equations,
which depend on the matter contents and interactions among them. Assuming only the SM particles, 
the SM gauge coupling constants come close to each other as the renormalization scale increases.
If we take a close look at the running, however,  they actually fail to unify 
unless rather large threshold corrections are introduced.
The gauge coupling unification is realized with a greater accuracy in various extensions
of the SM, such as supersymmetry \cite{susygut1,Ellis:1990wk,Amaldi:1991cn,Langacker:1991an,Giunti:1991ta}, introduction of incomplete multiplets 
 (see e.g. Refs.\cite{Murayama:1991ah, Bachas:1995yt}), etc. One simple resolution is to introduce unbroken hidden
U(1)$_H$ gauge symmetry with a large kinetic mixing $\chi$ with U(1)$_Y$~\cite{Redondo:2008zf}; the kinetic mixing with 
unbroken hidden U(1)$_H$ modifies the normalization of the hypercharge gauge coupling
in the high energy, thereby improving the gauge coupling unification. 
In this paper we focus on this simple resolution and argue that GUT with a hidden photon naturally leads to minicharged dark matter.
%

In Ref.~\cite{Takahashi:2016iph},  the two of the present authors (F.T.  and N.Y.), together 
with M.~Yamada, recently studied the gauge coupling unification with unbroken hidden U(1)$_H$
by solving the RG equations at one-loop level, including the effect of extra matter fields charged
under U(1)$_H$,  and discussed a possible origin of the required large kinetic mixing as well as
phenomenological and cosmological implications of the extra matter fields.
Those hidden matters are stable and contribute to dark matter. In particular, they 
acquire fractional electric charge through the large kinetic mixing, and such fractionally charged
stable matter has been searched for by many experiments \cite{Marinelli:1982dg, Davidson:1993sj, Davidson:2000hf, Lee:2002sa, Dubovsky:2003yn, Kim:2007zzs, McDermott:2010pa, Essig:2013lka, Dolgov:2013una, Moore:2014yba,Vogel:2013raa, Kamada:2016qjo}.

In this paper we study the GUT with a hidden photon in a greater detail and argue that minicharged dark matter
is its natural outcome. First of all we
 refine the analysis of Ref.~\cite{Takahashi:2016iph} by solving the RG equations
at two-loop level, and determine the GUT unification scale as well as the required size of the kinetic mixing precisely. 
The GUT unification scale turns out to be about $10^{16.5}$\,GeV which is high enough
to suppress the proton decay rate, and the required 
kinetic mixing is $\chi \simeq 0.37$ at the scale of the $Z$-boson mass.
Secondly, we find that the unification is almost determined by the kinetic mixing, but it is
rather insensitive to the size of the hidden gauge coupling or the presence of 
vector-like matter fields charged under U(1)$_H$ and/or SU(5).
As a consequence of the kinetic mixing, the hidden matter fields carry a non-zero electric charge, and they become minicharged dark matter if the hidden gauge coupling is sufficiently small.
 Thus the minicharged dark matter is a natural outcome of the GUT with a hidden photon. 
We will give concrete examples of such minicharged dark matter.\footnote{
See  e.g. Refs.~\cite{Cline:2012is,Foot:2014uba,Foot:2014osa,Foot:2016wvj,
Cardoso:2016olt} for recent works on minicharged dark matter.
The minicharged dark matter is often considered in a context of the mirror sector;
see Ref.~\cite{Foot:2014mia} for a comprehensive review on mirror dark matter.
%
}

The rest of this paper is organized as follows. In the next section we explain how the
gauge coupling unification is improved by adding U(1)$_H$, and show the results
of solving RG equations at two-loop level. In Sec.~\ref{sec:3} we discuss implications of
the hidden matter fields for minicharged dark matter. The last section is devoted for
discussion and conclusions.

\section{Gauge Coupling Unification with Hidden Photon}

\subsection{Preliminaries}
One way to improve unification of the SM gauge couplings is to modify the normalization of 
the U(1)$_Y$ gauge coupling at high energy scales. This can be realized by introducing
unbroken hidden gauge symmetry U(1)$_H$, which has a large kinetic mixing with U(1)$_Y$~\cite{Holdom:1985ag}. 
The relevant kinetic terms of the hypercharge and hidden gauge fields, $A_{\mu}'$ and $A_{H \mu}'$,  are given by
\begin{eqnarray}
\mathcal{L} = - \frac{1}{4} {F}_{\mu \nu}' F'^{\mu \nu} - \frac{1}{4} {F}'_{H \mu \nu} {F}_H'^{\mu \nu} - \frac{\chi}{2} F_{H \mu \nu}' F'^{\mu \nu} , \label{eq:org}
\end{eqnarray}
where $F_{\mu\nu}'$  and $F_{H \mu\nu}'$ are  gauge field strengths  of U(1)$_Y$ and U(1)$_H$, respectively. 
In this basis which we call the original basis in the following, the gauge fields and field strengths 
are indicated with a prime symbol.
For later use, we also introduce pairs of vector-like fermions,
\begin{eqnarray}
\mathcal{L} \ni - \sum_i M_{\Psi} \bar \Psi_i {\Psi_i} ,
\end{eqnarray}
where $\Psi_i$ has a hypercharge of $Q_i$ and a U(1)$_H$ charge  of ${q_H}_i$. 
The gauge interaction terms of the matter field $\Psi_i$ are written as
\begin{eqnarray}
\bar \Psi_i \gamma_\mu (g_Y' Q_i A'^\mu + g_H {q_H}_i A_H'^\mu) \Psi_i,
\end{eqnarray}
where $g_Y'$ and $g_H$ are the gauge couplings in the original basis.
We assume that hypercharges of vector-like fermions are rational numbers in the original basis such that
they can be embedded into the SU(5) GUT multiplet.

The canonically normalized gauge fields,  $A_{\mu}$ and $A_{H \mu}$, are obtained by the following transformations:
\begin{eqnarray}
A_{\mu}' = \frac{A_\mu}{\sqrt{1-\chi^2}}, \ \ 
A_{H \mu}' = A_{H \mu} - \frac{\chi}{\sqrt{1-\chi^2}} A_\mu,
\end{eqnarray}
and then, the kinetic terms become canonical, $\mathcal{L} = - \frac{1}{4}F_{\mu\nu}F^{\mu\nu} - \frac{1}{4}F_{H \mu\nu}F_H^{\mu\nu}$. In this canonical basis, gauge interaction terms of the matter field are written as
\begin{eqnarray}
\bar \Psi_i \gamma_\mu (g_Y' Q_i A'^\mu + g_H {q_H}_i A_H'^\mu) \Psi_i
=  \bar \Psi_i \gamma_\mu [(g_Y Q_i + g_{\rm mix} {q_H}_i )A^\mu + g_H {q_H}_i A_H^\mu ] \Psi_i,
\end{eqnarray}
where 
\begin{eqnarray}
g_Y = \frac{g_Y'}{\sqrt{1-\chi^2}}, \ \  g_{\rm mix} = - \frac{g_H \chi}{\sqrt{1-\chi^2}}.
\end{eqnarray}
Here, $g_Y$ is the gauge couplings of U(1)$_Y$, and $g_H$ remains unchanged by the transformation from the
original basis to the canonical one.
One can see that the field $\Psi_i$ now acquires a fractional hypercharge, $ g_{\rm mix} {q_H}_i/g_Y$, 
which is a renormalization scale dependent quantity~\cite{Holdom:1985ag,Glashow:1985ud,Carlson:1987si}. 
The U(1)$_Y$ coupling with a prime, $g_Y'$, is the gauge coupling in the original basis (see Eq.(\ref{eq:org})), and is smaller by $\sqrt{1-\chi^2}$ compared to $g_Y$. Thus, the kinetic mixing with unbroken U(1)$_H$ modifies the normalization of the hypercharge
coupling constant, and the unification of the gauge couplings can be improved by choosing $\chi$ so that
 $\sqrt{\frac{5}{3}}g_Y'$ at the GUT scale is equal to the unified gauge coupling determined by the running of $g_2$ and $g_3$.\footnote{
We emphasize here that the kinetic mixing with hypercharge is only able to suppress the gauge coupling $g_Y'$
compared to $g_Y$. On the other hand, introducing extra matter fields with hypercharge has the opposite effect on $g_Y$ and does not
improve the unification.
 }
In other words, the gauge coupling unification is realized in the original basis where the kinetic mixing is manifest. We shall return to the origin of such kinetic mixing later in this section.

In the canonical basis, $\Psi_i$ has the fractional U(1)$_Y$ charge and its effect is captured by
 the beta-functions of the gauge couplings. 
The actual calculations to be given in the next subsection are based on the two-loop RG equations,
but let us give the one-loop RG equations below to get the feeling of how the gauge couplings evolve.

 The one-loop beta-functions of the gauge couplings in the canonical basis are given by~\cite{Babu:1996vt}
\begin{eqnarray}
\frac{d g_Y}{dt} &=&  \frac{1}{16\pi^2} (b_Y g_Y^3 + b_H g_Y g_{\rm mix}^2 + 2b_{\rm mix} g_Y^2 g_{\rm mix} ), \nonumber \\
\frac{d g_H}{dt} &=&  \frac{1}{16\pi^2} b_H g_H^3 ,\nonumber \\
\frac{d g_{\rm mix}}{dt} &=&  \frac{1}{16\pi^2} (b_Y g_{\rm mix} g_Y^2 + 2 b_H g_{\rm mix} g_H^2  + b_H g_{\rm mix}^3 + 2 b_{\rm mix} g_Y g_H^2 + 2 b_{\rm mix} g_Y g_{\rm mix}^2 ), \label{eq:rge1}
\end{eqnarray}
where $t = \ln \mu_R$ ($\mu_R$ is a renormalization scale) and 
\begin{eqnarray}
b_Y = \frac{41}{6} + \frac{4}{3} \sum_i Q_i^2, \ \ 
b_H = \frac{4}{3} \sum_i {q_H}_i^2, \ \ 
b_{\rm mix} = \frac{4}{3} \sum_i Q_i {q_H}_i \,.
\end{eqnarray}
On the other hand, the beta-functions of the gauge couplings and the kinetic mixing parameter in the 
original basis take a surprisingly simple form. By using Eq.\,(\ref{eq:rge1}), 
the beta-functions in the original basis are written as
\begin{eqnarray}
\frac{d g_Y'}{dt} &=&  \frac{1}{16\pi^2} b_Y g_Y'^3 \, ,\nonumber \\
\frac{d  g_H}{dt} &=&  \frac{1}{16\pi^2} b_H g_H^3 \, , \nonumber \\
\frac{d \chi}{dt} &=&  \frac{1}{16\pi^2} \left[\chi (b_Y g_Y'^2+ b_H g_H^2)
-2 b_{\rm mix} g_Y' g_H \right].
\end{eqnarray}
Note that the RG running of $g_Y'$ does not depend on $g_H$ nor $\chi$ at the one-loop level,
although its normalization is fixed by $\chi$.
At the two-loop level,
 this property does not hold and the RG running of $g_Y'$
depends on $g_H$ and $\chi$ (see in Appendix \ref{appB}).
However, the dependence is still weak due to the loop suppression factor, 
and we have  confirmed this numerically.  Therefore, because of rather weak dependence on
$g_H$ and $\chi$, 
the gauge coupling unification is essentially determined by the size of $\chi$.

\subsection{Numerical results of solving RG equations}
Now we study the RG runnings of the gauge couplings using two-loop beta-functions,\footnote{
Apart from the gauge couplings, we only take into account the top-Yukawa coupling and Higgs quartic coupling. In the numerical analysis, we use one-loop RG equations for these couplings.
}
in order to see if the SM gauge couplings unify at the high-energy scale. 
The two-loop beta-functions are obtained by utilizing {\tt PyR@TE 2} package~\cite{Lyonnet:2013dna, Lyonnet:2016xiz},
and the numerical results shown in this section are based on solving the RG equations at two-loop order unless otherwise
stated.

\begin{figure}[!t]
\begin{center}
\includegraphics[bb=0 0 200 200, scale=1.0]{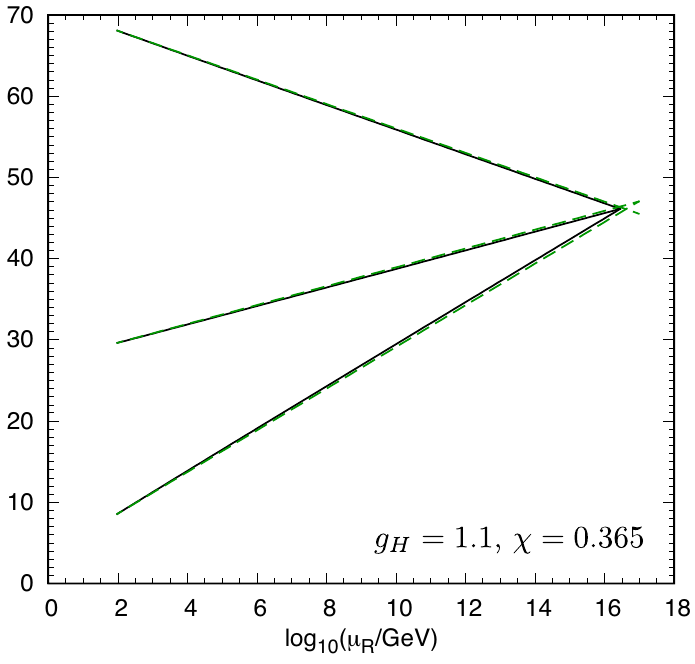}
\includegraphics[bb=0 0 200 200, scale=1.0]{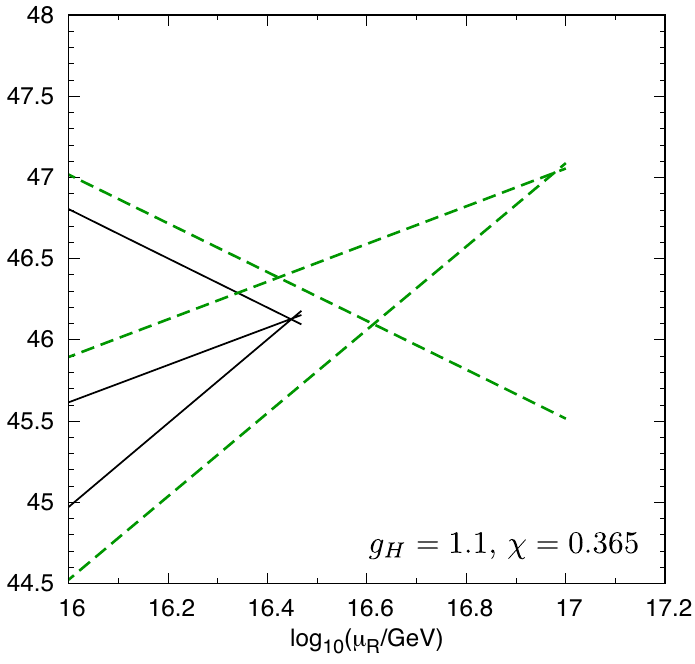}
\caption{
The RG runnings of gauge couplings.  
We take $\chi=0.365$ at $m_Z$. (In the case with hidden matter fields,
we further set  $g_H=1.1$)
The black solid (green dashed) lines show the result using two-loop (one-loop) RG equations. 
On the right panel, the region of $\mu_R$ around $10^{16}{\mathchar`-}10^{17}$ GeV is zoomed.
Here, $\alpha_s(m_Z)=0.1185$ and $m_t({\rm pole})=173.34$\,GeV.
}
\label{fig:1}
\end{center}
\end{figure}

Let us first study the case without extra matter fields by solving the RG equations at the two-loop order. 
The one-loop analysis in this case was studied
in Ref.~\cite{Redondo:2008zf}. The beta-functions of the gauge couplings at the one-loop level are given by
\begin{eqnarray}
\frac{d g_Y'}{dt} &=&  \frac{1}{16\pi^2} \left(\frac{41}{6}\right) g_Y'^3, \nonumber \\
\frac{d g_2}{dt} &=&  \frac{1}{16\pi^2} \left(- \frac{19}{6}\right) g_2^3, \nonumber \\
\frac{d g_3}{dt} &=&  \frac{1}{16\pi^2} \left(-7\right) g_3^3, \nonumber 
\end{eqnarray}
where $g_2$ and $g_3$ are the gauge couplings of SU(2)$_L$ and SU(3)$_C$, respectively. 
The hidden gauge coupling does not run  in this case. 
In Fig.\,\ref{fig:1}, we plot the RG running of $\alpha_1'^{-1}$, $\alpha_2^{-1}$ and $\alpha_3^{-1}$, where $\alpha_1' = \frac{5}{3}  g_Y'^2/(4\pi)$, $\alpha_{2} = g_{2}^2/(4\pi)$ and $\alpha_3=g_3^2/(4\pi)$. 
The black solid (green dashed) lines show the result computed using two-loop (one-loop) beta-functions. 
We take $\chi=0.365$ at the scale of the $Z$ boson mass, $m_Z$. (The value of $g_H$ shown in the figure is
for the next case we study below.)
As one can see, the difference between the one-loop and two-loop results are not large, but
the expected unification scale with the two-loop calculation is around $10^{16.5}$\,GeV,
which is slightly smaller than that with the one-loop calculation.
%

The above case without any extra matter fields captures the essence of how the kinetic mixing between hypercharge
and hidden U(1)$_H$ improves the unification. However, as emphasized in Ref.~\cite{Redondo:2008zf}, there is no
phenomenological implications for low-energy physics (except for suppressed proton decay rates), since the massless hidden photon is decoupled from
the SM sector, and the hidden U(1)$_H$ simply changes the normalization
of the U(1)$_Y$ gauge coupling. In particular, from the low-energy physics point of view, there is no way to
determine the correct basis at the high energy except for requiring the successful gauge coupling unification, 
since any basis appears to be on an equal footing.

Next we introduce a vector-like fermion, which is charged only under U(1)$_H$,
\begin{eqnarray}
\mathcal{L} \ni - M_{0} \bar \Psi_0 \Psi_0,
\end{eqnarray}
with $(Q_0, {q_H}_0)=(0, 1)$. 
The beta-functions of the SM gauge couplings at the one-loop level are 
same as above and the beta-function of the U(1)$_H$ is 
given by
\begin{eqnarray}
\frac{d g_H}{dt} &=&  \frac{1}{16\pi^2} \left(\frac{4}{3} \right) g_H^3.
\end{eqnarray}
In the numerical calculations we set $M_0 = 1$\,TeV, $g_H = 1.1$ and $\chi = 0.365$. In this case, the results have turned out to be
essentially same as Fig.~\ref{fig:1}.
We have confirmed that the RG runnings of the gauge couplings as well as the unification scale are rather insensitive to 
$g_H$ even at the two-loop level, by varying $g_H$ from $10^{-4}$ to $1.1$.
Note that, in this scenario with hidden particles, the basis where the gauge coupling unification occurs 
is manifest, because the hyperchages of hidden matter fields need to be quantized (including zero) so that they are 
consistent with the SU(5) GUT gauge group.

\begin{figure}[!t]
\begin{center}
\includegraphics[bb=0 0 200 200, scale=1.0]{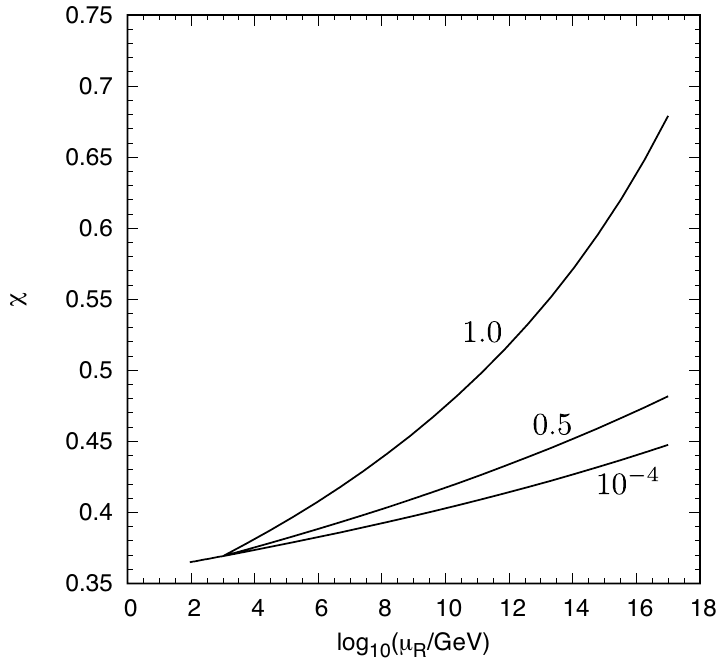}
\caption{
The RG running of the kinetic mixing for $g_H = 10^{-4}$, $0.5$ and $1.0$  at $m_Z$
from bottom to top.
}
\label{fig:2}
\end{center}
\end{figure}

Although the RG running of $g_Y'$ is almost insensitive to $g_H$ even at the two-loop level, 
the running of $\chi$ depends sensitively on the size of $g_H$. We show RG running of $\chi$ for different values of $g_H$ in Fig.\,\ref{fig:2}. For a large $g_H$ as $1.0$, $\chi$ at $10^{17}$\,GeV becomes large as $0.7$, while if $g_H$ is smaller than $0.5$, $\chi$ at $10^{17}$\,GeV remains around $0.45$\,-\,$0.5$.

\vspace{10pt}
Next we consider the case where there are $N_{\rm bi}$ pairs of bi-charged vector-like fermions:
\begin{eqnarray}
\mathcal{L} = - M_V \sum_{i=1}^{N_{\rm bi}} ( \bar{\Psi}_{L,i}  {\Psi_{L,i}}  
+ {\bar \Psi}_{\bar D,i} \Psi_{\bar  D,i}), \label{eq:bicharged}
\end{eqnarray}
where $\Psi_{L,i}$ ($\Psi_{\bar D,i}$) is {\bf 2} of SU(2)$_L$ ($\bar {\bf 3}$ of SU(3)$_C$);
$(Q_{L,i}, {q_H}_{L,i})=(-1/2, 1)$ and $(Q_{\bar D,i}, {q_H}_{\bar D,i})=(1/3, 1)$. Here, $\Psi_{L,i}$ and $\Psi_{\bar D,i}$ form a complete SU(5) multiplet. In this case, $b_{\rm mix}$ vanishes. (See Appendix \ref{appA} for a case where
 $\Psi_{L,i}$ and $\Psi_{\bar D_i}$  have different $q_H$ and do not form a complet multiplet.)
The one-loop beta-functions of the gauge couplings are
\begin{eqnarray}
\frac{d g_Y'}{dt} &=&  \frac{1}{16\pi^2} \left(\frac{41}{6}  + \frac{10}{9} N_{\rm bi} \right) g_Y'^3, \nonumber \\
\frac{d g_2}{dt} &=&  \frac{1}{16\pi^2}  \left(-\frac{19}{6} + \frac{2}{3} N_{\rm bi} \right) g_2^3, \nonumber \\
\frac{d g_3}{dt} &=&  \frac{1}{16\pi^2} \left(-7 + \frac{2}{3} N_{\rm bi} \right) g_3^3, \nonumber \\
\frac{d g_H}{dt} &=&  \frac{1}{16\pi^2} \left(\frac{20}{3} N_{\rm bi}\right) \, g_H^3.  
\end{eqnarray}

\begin{figure}[!t]
\begin{center}
\includegraphics[bb=0 0 200 200, scale=1.0]{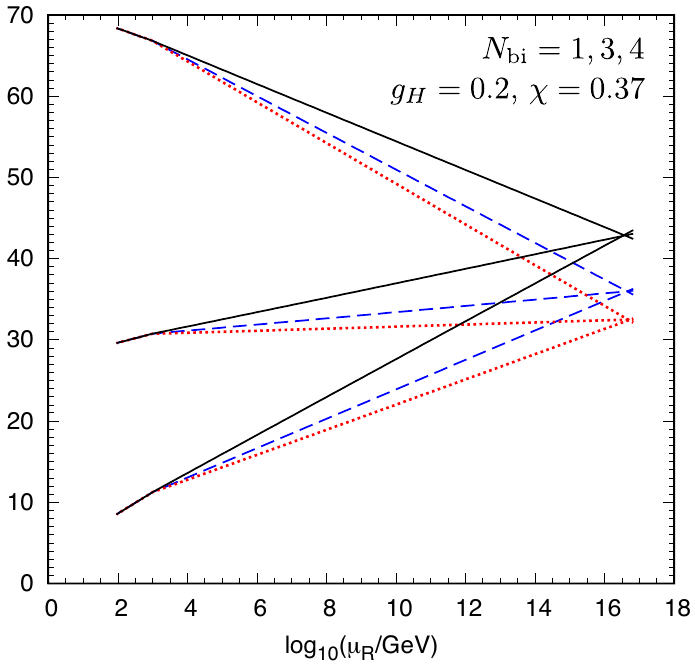}
\includegraphics[bb=0 0 200 200, scale=1.0]{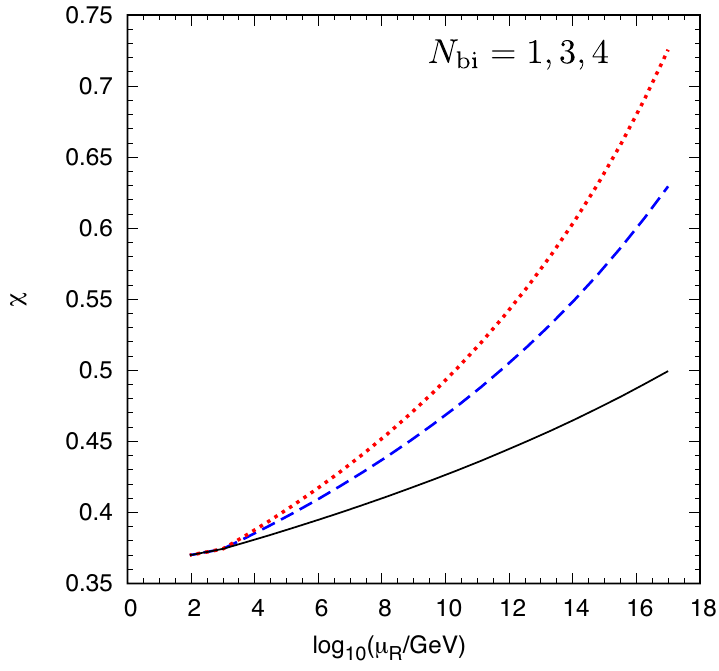}
\caption{
The RG runnings of the gauge couplings (left) and mixing (right) with $N_{\rm bi}$ bi-charged fields.
The cases of $N_{\rm bi} = 1, 3,$ and $4$ are represented by black solid, blue dashed, and red
dotted lines, respectively. 
We take $g_H=0.2$ and $\chi=0.37$ at $m_Z$. The mass of the bi-charged field, $M_V$,  is set to be 1\,TeV.
}
\label{fig:3}
\end{center}
\end{figure}

\begin{figure}[!t]
\begin{center}
\includegraphics[bb=0 0 200 200, scale=1.0]{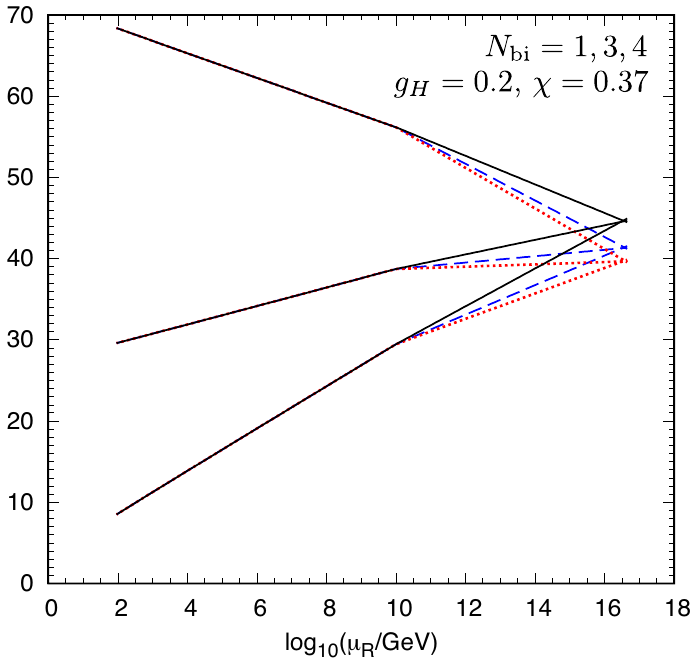}
\includegraphics[bb=0 0 200 200, scale=1.0]{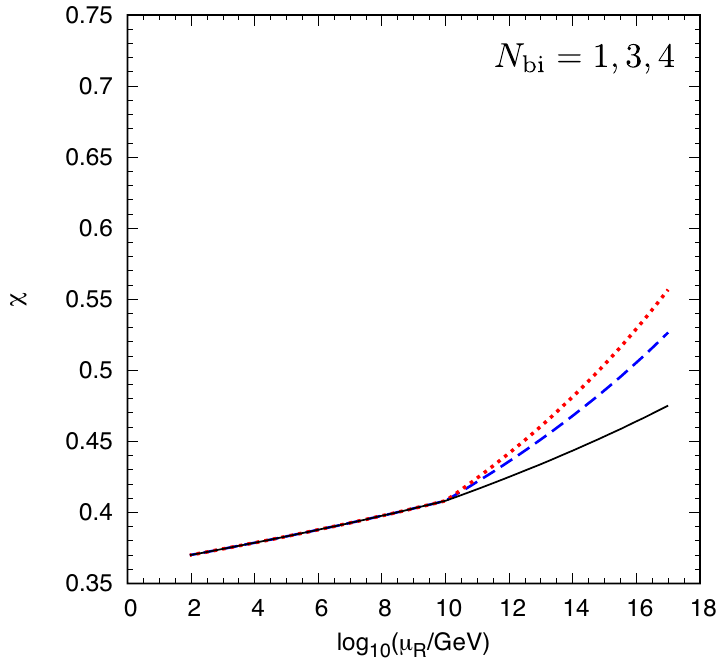}
\caption{
Same as Fig.~\ref{fig:3} but for $M_{V}=10^{10}$ GeV.
}
\label{fig:4}
\end{center}
\end{figure}

In Fig.\,\ref{fig:3} and \ref{fig:4}, we show the RG runnings of the gauge couplings and the 
mixing parameter for $N_{\rm bi}=1,3,4$, based on the two-loop beta-functions.  
We set $g_H$ and $\chi$ at $m_Z$ as $g_H=0.2$ and $\chi=0.37$.
In Fig.\,\ref{fig:3} (Fig.\,\ref{fig:4}), we take $M_V=1$\,TeV ($10^{10}$ GeV). 
One can see that the unification scale does not depend on $N_{\rm bi}$ nor $M_V$. Also,  the required value of 
$\chi$ at $m_Z$ for the unification remains almost intact for different choices of $N_{\rm bi}$ and $M_V$. 
On the other hand, $\chi$ at the high-energy scale (e.g. $10^{17}$ GeV) is sensitive to the change of $N_{\rm bi}$ and $M_V$. 

Thus, we have found that, once $\chi$($m_Z$) is fixed to be around 0.37, 
we can freely choose $N_{\rm bi}$, $M_V$ and $g_H$ without affecting the gauge coupling unification.
In particular the unification is realized even with a tiny hidden gauge coupling.
This feature is suitable to identify the hidden matter as  dark matter, 
which can have a tiny electric charge (in the canonical basis).
Interestingly, the stability of the dark matter is ensured by the unbroken hidden U(1)$_H$ gauge symmetry.

\subsection{Origin of the kinetic mixing}
Finally, let us comment on the origin of the kinetic mixing $\chi$ and a possible modification of the unified gauge coupling. 
We have seen that a relatively large $\chi \sim 0.37$ is necessary to improve the gauge coupling unification.
Such a large kinetic mixing may be generated via an operator,
\begin{eqnarray}
\mathcal{L} \ni \frac{k_0}{M_*} {\rm Tr} (\left<\Sigma_{24}\right> F_{5 \mu \nu}) F_H^{\mu \nu},
\end{eqnarray}
where $\Sigma_{24}$ is a GUT breaking Higgs and $\left<\Sigma_{24}\right> = (2,2,2,-3,-3) v_{24}/(2\sqrt{15})$ with $v_{24} \sim 10^{17}$\,GeV; $M_*$ is a cut-off scale.
Then, the kinetic mixing is given by
\begin{eqnarray}
\chi = - k_0 \frac{v_{24}}{M_*}.
\end{eqnarray}
For  $M_*\sim 10^{18}$\,GeV, the induced kinetic mixing is naturally of ${\cal O}(0.1)$,
as required for the successful unification.

We also note that there could also exist an operator, 
\begin{eqnarray}
\mathcal{L} \ni \frac{k_1}{2 M_*} {\rm Tr} (\left<\Sigma_{24}\right> F_{5 \mu \nu} F_5^{\mu \nu}),
\end{eqnarray}
which modifies the unified gauge coupling. It leads to
the following deviations,
\begin{eqnarray}
\frac{\Delta \alpha_1^{-1}}{ \alpha_{5}^{-1}} = \sqrt{\frac{1}{60}} \frac{k_1}{k_0} \chi, \ 
\frac{\Delta \alpha_2^{-1}}{ \alpha_{5}^{-1}} = \sqrt{\frac{3}{20}} \frac{k_1}{k_0} \chi, \ 
\frac{\Delta \alpha_3^{-1}}{\alpha_{5}^{-1}} = -\sqrt{\frac{1}{15}} \frac{k_1}{k_0} \chi,
\end{eqnarray}
where $\alpha_5$ is the squared of the unified coupling divided by 4$\pi$. 
Therefore, if $k_1/k_0 \lesssim 6$, the deviation, $(\Delta\alpha_2^{-1}-\Delta\alpha_3^{-1})/\alpha_5^{-1}$, is within 5\% level while obtaining the large kinetic mixing of $0.5$.

\section{Minicharged Dark Matter}
\label{sec:3}

Let us suppose that there is a field, $\Psi_H$, which is charged only under U(1)$_H$ in the original basis.
Then, the field acquires an electric charge $q_e$ proportional to the hidden gauge coupling $g_H$, and so,
if $g_H$ is tiny, it becomes a minicharged particle. Since the stability of $\Psi_H$ is ensured by the U(1)$_H$, 
such minicharged particle can be a good candidate for dark matter.

In order to account for dark matter, such minicharged particles must be somehow produced in the early Universe.
As long as we assume thermal production through electromagnetic interactions, 
however, they can not be a dominant component of the dark matter. This is because
 the astrophysical constraints~\cite{Davidson:1993sj, Davidson:2000hf, Dubovsky:2003yn, McDermott:2010pa, Essig:2013lka, Dolgov:2013una, Vogel:2013raa, Kamada:2016qjo} and the direct detection constraint~\cite{DelNobile:2015bqo}\,\footnote{
If the minicharged particle is not a dominant component of dark matter,  the constraint from the direct detection experiment may be avoided in the following range of $q_e$~\cite{McDermott:2010pa}:
\begin{eqnarray}
5.4 \times 10^{-10} \left(\frac{m_{X,Y}}{1\,{\rm TeV}}\right) \lesssim |q_e| \lesssim 1.1 \times 10^{-2} \left(\frac{m_{X,Y}}{1\,{\rm TeV}}\right)^{1/2},
\end{eqnarray}
since the minicharged particle may be evacuated from the Galactic disk by the supernova shock waves and Galactic magnetic fields~\cite{Chuzhoy:2008zy}.
} on the minicharged particle exclude the region where the thermal relic abundance is consistent with the observed dark matter abundance. 
To obtain the correct thermal relic abundance without running afoul of astrophysical and direct 
detection constraints, one needs to include additional interactions between the minicharged particles and
the SM sector.

To be concrete, we consider the following Higgs portal interactions~\cite{Davoudiasl:2004be, Patt:2006fw,Kim:2006af, Barger:2007im, Kim:2008pp}:
\begin{eqnarray}
\mathcal{L} \ni -m_X^2 |X|^2 + \lambda_X |H|^2 |X|^2  \ \ {\rm or} \ \   
-m_Y \bar \Psi_Y \Psi_Y + \frac{\lambda_Y}{\Lambda_Y} |H|^2 \bar \Psi_Y \left(\frac{1-\gamma_5}{2}\right) \Psi_Y + h.c., \label{eq:higgs_portal}
\end{eqnarray}
where $H$ is a SM Higgs doublet, $\Lambda_Y$ is some mass scale, 
$X$ and $\Psi_Y$ are respectively a complex scalar and a Dirac fermion with a unit U(1)$_H$ charge,
and $m_X$ and $m_Y$ are their masses. 
In the canonical basis, $X$ and $\Psi_Y$ have an electric charges of 
\begin{eqnarray}
q_e = -\frac{ \chi }{\sqrt{1-\chi^2}} \frac{g_H}{g_Y}.
\label{eq:qe}
\end{eqnarray}
If $m_{X,Y} \gg \left<H^0\right>$,
the correct abundance is obtained for $\lambda_X \simeq 0.6$ and $m_X \simeq 1\,{\rm TeV}$ or $\lambda_Y/\Lambda_Y \simeq 10^{-3.5}\,{\mathchar`-}\,10^{-3.25}$\,GeV$^{-1}$ and $m_Y \simeq\,$1\,-\,10\,TeV~\cite{Beniwal:2015sdl}, avoiding the constraint from the LUX experiment~\cite{Akerib:2016vxi}. As long as the electric charge $|q_e|$ is sufficiently small, this allowed region does not depend on $|q_e|$. Therefore, the minicharged dark matter can explain the observed dark matter if it has a Higgs portal coupling. 

The minicharge of dark matter is constrained by various observations.
First, if the minicharged dark matter is tightly coupled to the photon-baryon plasma 
during  recombination, the power spectrum of the Cosmic Microwave Background anisotropies is modified and it 
becomes inconsistent with observations~\cite{Dubovsky:2003yn, McDermott:2010pa, Dolgov:2013una,Kamada:2016qjo}. 
Requiring that the dark matter is completely decoupled from the plasma at the recombination epoch, 
an upper-bound on $q_e$ is obtained as~\cite{McDermott:2010pa}
\begin{eqnarray}
|q_e| \lesssim 10^{-4} \left(\frac{m_{X,Y}}{1\,{\rm TeV}} \right)^{1/2},
\end{eqnarray}
which requires $g_H \lesssim 10^{-4}$.
Direct detection experiments give even tighter constraints.
From the LUX experiment~\cite{Akerib:2016vxi}, the constraint on $q_e$ is~\cite{DelNobile:2015bqo}
\begin{eqnarray}
|q_e| \lesssim 3.6 \times 10^{-10} \left(\frac{m_{X,Y}}{1\,{\rm TeV}}\right)^{1/2}.
\end{eqnarray}
%
%
%
Note that this LUX constraint can be applied to very heavy minicharged dark matter.

Finally, let us comment on an alternative production of the minicharged dark matter. 
The minicharged particle may have an interaction with a heavy particle (e.g. inflaton),
and the correct relic abundance may be obtained by non-thermal productions via the interaction.
For instance, if the inflaton has a quartic coupling with $X$, and oscillates about the origin where the $X$ becomes
(almost) massless, a preheating process could take place. For certain coupling and the mass of $X$, the right abundance
of $X$ can be generated. One way to suppress the overproduction of $X$ is consider a minicharged dark matter of
a heavy mass.
Such scenario is also consistent with the phenomenological constraints as long as the electric charge is small enough.

\section{Discussion and Conclusions}
\label{sec:4}

We have investigated the gauge coupling unification in the presence of
an unbroken hidden U(1)$_H$ symmetry, which mixes with the U(1)$_Y$ of the SM gauge group. 
By solving the two-loop RG equations, 
we have found the gauge coupling unification is achieved with a better accuracy if the size of the kinetic mixing is 
$\chi \simeq 0.37$ at the Z boson mass scale. 
The unification scale is around $10^{16.5}$\,GeV, which is large enough to avoid the rapid proton decay.
Interestingly, the unification behavior is essentially determined by the kinetic mixing parameter only, 
and it is rather insensitive to the size of the hidden gauge coupling or the presence of the vector-like fermions 
charged under U(1)$_H$ and/or SU(5).
This implies that the vector-like masses can be arbitrarily light (or heavy) without affecting the gauge coupling unification.  

The above findings imply that the Peccei-Quinn mechanism to the strong CP-problem~\cite{Peccei:1977hh,Weinberg:1977ma,Wilczek:1977pj}  can be easily embedded into our setup: if the vector-like fermion of the SU(5) complete multiplet
is coupled to a Peccei-Quinn scalar, it induces the required color anomaly~\cite{Kim:1979if,Shifman:1979if}. 
The gauge coupling unification is preserved irrespective of the axion decay constant, which is
typically around the intermediate scale. Interestingly, if the vector-like fermion has a hidden U(1) charge,
the axion is coupled to the hidden photon, and the axion will be a portal to the hidden photon.
In this case, the dark matter could be composed of both the QCD axion and the minicharged dark matter.

We have shown that the U(1)$_H$ gauge coupling can be arbitrarily small while keeping the successful unification. 
In this case, a hidden particle charged under U(1)$_H$ has a tiny electric charge due to the kinetic mixing. 
The U(1)$_H$ charge ensures the stability of the particle; therefore, the minicharged particle is a natural 
candidate for dark matter. The minicharged dark matter can have a correct relic abundance through the Higgs portal interactions~\footnote{
Similarly, the minicharged dark matter is expected to have the correct relic abundance through the axion portal~\cite{Nomura:2008ru}. The benefit of this case is that the nonrenormalizable interaction (Eq.(\ref{eq:higgs_portal})) is not needed for the fermionic dark matter.
}  while avoiding known phenomenological constraints. 
Thus, the minicharged dark matter naturally arises from GUT with a hidden photon.

\begin{figure}[!t]
\begin{center}
\includegraphics[bb=0 0 200 200, scale=1.0]{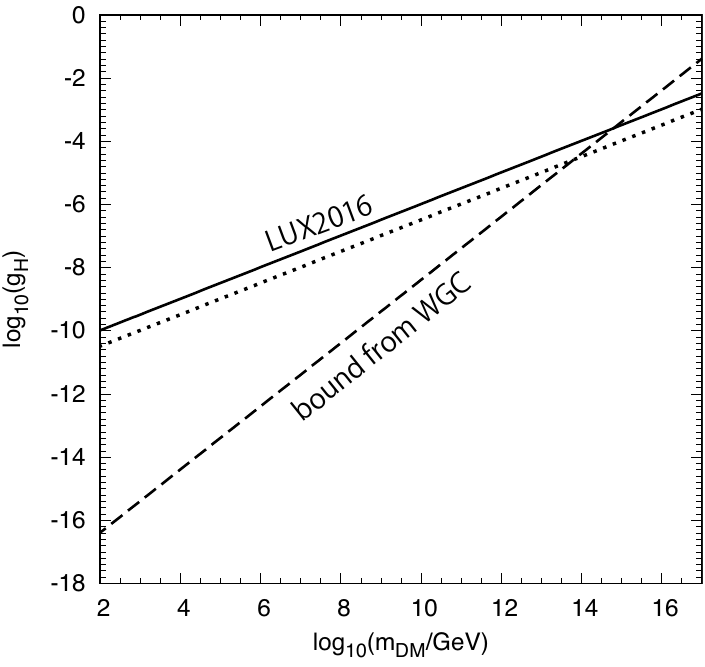}
\caption{
The lower bound on $g_H$ from Weak Gravity Conjecture is shown. 
The region below the dashed line is conflict with WGC. We also show the upper bound on $g_H$ from LUX 2016 (solid) 
and the projected sensitivity of XENON 1T (dotted).
}
\label{fig:wgc}
\end{center}
\end{figure}

Finally, let us comment on the possible lower bound on the hidden gauge coupling. 
From the Weak Gravity Conjecture (WGC)~\cite{wgc}, which claims that the gravity is the weakest force, 
the hidden gauge coupling must satisfy the constraint, $g_H(m_{X,Y}) > m_{X,Y}/M_{PL}$,
where $M_{PL}$ is the Planck mass.\footnote{
Here, we adopt a version of the conjecture that the mass of the lightest charged particle $m_{l}$ should satisfy $m_{l} < g_H(m_l) M_{PL}$.
This constraint is shown in Fig.~\ref{fig:wgc}, together with the upper bound from the LUX experiment. The WGC also claims that the cut-off scale of U(1)$_X$, $\Lambda$, is smaller than $g_H (\Lambda) M_{PL}$. 
Requiring that $\Lambda$ be larger than the GUT scale, $g_H$ needs to satisfy $g_H (M_{\rm GUT}) > M_{\rm GUT}/M_{PL}$ with $M_{\rm GUT} \sim 10^{17}\,$GeV. This condition predicts many hidden particles with masses between $m_{X,Y}$ and $M_{\rm GUT}$, leading to large enough $g_{H}(M_{\rm GUT})$ of $\sim 10^{-2}$.
}

\section*{Acknowledgments}

F.T. thanks K. Kohri for useful discussion on the cosmological effects of 
charged massive particles.  
This work is supported by 
Tohoku University Division for Interdisciplinary Advanced Research and Education (R.D.);
JSPS KAKENHI Grant Numbers 15H05889 and 15K21733 (F.T. and N.Y.);
JSPS KAKENHI Grant Numbers  26247042 and 26287039 (F.T.),
and by World Premier International Research Center Initiative (WPI Initiative), MEXT, Japan (F.T.).

\appendix

\section{A case with flipped hidden charges}
\label{appA}

\begin{figure}[!t]
\begin{center}
\includegraphics[bb=0 0 200 200, scale=1.0]{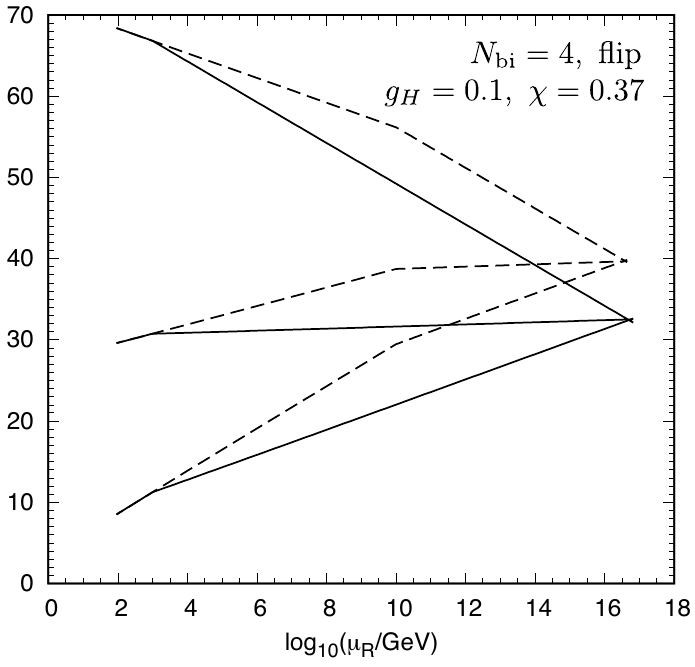}
\caption{
The RG runnings of the gauge couplings for $M_V=1$\,TeV (solid) and $10^{10}$\,GeV (dashed) with flipped hidden charges. Here, $g_H=0.1$ and $\chi=0.37$ at $m_Z$.
}
\label{fig:flip}
\end{center}
\end{figure}

In Eq.\,(\ref{eq:bicharged}), $\Psi_{L,i}$ and $\Psi_{\bar D,i}$  form a complete SU(5) multiplet, and $b_{\rm mix}$ vanishes. Here, we consider a difference case: $\Psi_{L,i}$ and $\Psi_{\bar D,i}$ have flipped charges as  $(Q_{L,i}, {q_H}_{L,i})=(-1/2, 1)$ and $(Q_{\bar D,i}, {q_H}_{\bar D,i})=(1/3, -1)$ leading to non-vanishing $b_{\rm mix}$. In Fig.~\ref{fig:flip}, we show the results for $M_V=10^3$\,GeV (solid line) and $10^{10}$\,GeV (dashed line). Again, the unification does occur with $\chi (m_Z) \simeq 0.37$.

\section{Two loop RG equations}
\label{appB}
Here we give the relevant RG equations at two-loop level following to Ref.~\cite{Luo:2002iq}. 
\subsection{Hidden vector-like fermion}
In the case with $N_{\rm vec}$ hidden vector-like fermions, the two-loop RGEs are given as follows. Here, $q_{H,i}$ denotes $U(1)_H$ charge of $\Psi_{0,i}$.
\begin{align}
\frac{d g_Y'}{dt}&=\frac{g_Y'^3}{16\pi^2}\left(\frac{41}{6}\right) \nonumber \\
&+\frac{g_Y'^3}{(16\pi^2)^2}\left[\left(\frac{199}{18}\right)\frac{g_Y'^2}{1-\chi^2} +\left(\frac{9}{2}\right)g_2^2+\left(\frac{44}{3}\right)g_3^2-\frac{17}{6}y_t^2\right] \nonumber \\
\frac{d g_H}{dt}&=\frac{g_H^3}{16\pi^2}\left(\frac{4}{3}\sum_{i=1}^{N_{\rm vec}}q_{Hi}^2\right) \nonumber \\
&+\frac{g_H^3}{(16\pi^2)^2}\left[\left(4\sum_{i=1}^{N_{\rm vec}}q_{Hi}^4\right)\frac{g_H^2}{1-\chi^2}\right] \nonumber \\
\frac{d \chi}{dt}&=\frac{\chi}{16\pi^2}\left[\left(\frac{41}{6}\right)g_Y'^2+\left(\frac{4}{3}\sum_{i=1}^{N_{\rm vec}}q_{Hi}^2\right)g_H^2\right] \nonumber \\
&+\frac{1}{(16\pi^2)^2}\left[\left(\frac{199}{18}\right)\frac{g_Y'^4\chi}{1-\chi^2}+\left(\frac{9}{2}\right)\chi g_Y'^2g_2^2+\left(\frac{44}{3}\right)\chi g_Y'^2g_3^2+\left(4\sum_{i=1}^{N_{\rm vec}}q_{Hi}^4\right)\frac{g_H^4\chi}{1-\chi^2}-\frac{17}{6}g_Y'^2y_t^2\chi \right] \nonumber \\
\frac{d g_2}{dt}&=\frac{g_2^3}{16\pi^2}\left(-\frac{19}{6}\right) \nonumber \\
&+\frac{g_2^3}{(16\pi^2)^2}\left[\left(\frac{3}{2}\right)\frac{g_Y'^2}{1-\chi^2}+\left(\frac{35}{6}\right)g_2^2+\left(12\right)g_3^2-\frac{3}{2}y_t^2\right] \nonumber \\
\frac{d g_3}{dt}&=\frac{g_3^3}{16\pi^2}\left(-7\right) \nonumber \\
&+\frac{g_3^3}{(16\pi^2)^2}\left[\left(\frac{11}{6}\right)\frac{g_Y'^2}{1-\chi^2}+\left(\frac{9}{2}\right)g_2^2+\left(-26\right)g_3^2-2y_t^2\right] \nonumber \\ 
\frac{d y_t}{dt}&=\frac{y_t}{16 \pi ^2} \left(-\frac{17}{12}\frac{g_Y'^2}{1-\chi2}-\frac{9}{4}g_2^2-8g_3^2+\frac{9}{2}y_t^2\right) \nonumber \\
\frac{d \lambda}{dt}& =\frac{1}{16\pi^2}\left(24 \lambda^{2} -3\lambda \frac{g_Y'^{2}}{1-\chi^2} -9\lambda g_{2}^{2}  +12 \lambda y_t^2+\frac{3}{8}  \frac{g_Y'^{4}}{(1-\chi^2)^2} +\frac{3}{4}  \frac{g_Y'^{2}}{1-\chi^2} g_{2}^{2} +\frac{9}{8} g_{2}^{4} -6y_t^4\right)\nonumber 
\end{align}

\subsection{Bi-charged vector-like fermion}
In the case with $N_{\rm bi}$ pairs of bi-charged vector-like fermions, the two-loop RGEs are given as follows. Here, we assume that $\Psi_{L,i}$ ($\Psi_{\bar D,i}$) is {\bf 2} of SU(2)$_L$ ($\bar {\bf 3}$ of SU(3)$_C$);
$(Q_{L,i}, {q_H}_{L,i})=(-1/2, q_{H,i})$ and $(Q_{\bar D,i}, {q_H}_{\bar D,i})=(1/3, q_{H,i})$.
\begin{align}
\frac{d g_Y'}{dt}&=\frac{g_Y'^3}{16\pi^2}\left(\frac{41}{6}+\frac{10}{9}N_{\rm bi}\right) \nonumber \\
&+\frac{g_Y'^3}{(16\pi^2)^2}\left[\left(\frac{199}{18}+\frac{35}{54}N_{\rm bi}\right)\frac{g_Y'^2}{1-\chi^2}+\left(\frac{10}{3}\sum_{i=1}^{N_{\rm bi}}q_{Hi}^2\right)\frac{g_H^2}{1-\chi^2}+\left(\frac{10}{9}\sum_{i=1}^{N_{\rm bi}}q_{Hi}\right)\frac{g_Y'g_H\chi}{1-\chi^2} \right. \nonumber \\
&\left.+\left(\frac{9}{2}+\frac{3}{2}N_{\rm bi}\right)g_2^2+\left(\frac{44}{3}+\frac{16}{9}N_{\rm bi}\right)g_3^2-\frac{17}{6}y_t^2\right] \nonumber \\
\frac{d g_H}{dt}&=\frac{g_H^3}{16\pi^2}\left(\frac{20}{3}\sum_{i=1}^{N_{\rm bi}}q_{Hi}^2\right) \nonumber \\
&+\frac{g_H^3}{(16\pi^2)^2}\left[\left(\frac{10}{3}\sum_{i=1}^{N_{\rm bi}}q_{Hi}^2\right)\frac{g_Y'^2}{1-\chi^2}+\left(20\sum_{i=1}^{N_{\rm bi}}q_{Hi}^4\right)\frac{g_H^2}{1-\chi^2}+\left(6\sum_{i=1}^{N_{\rm bi}}q_{Hi}^2\right)g_2^2+\left(16\sum_{i=1}^{N_{\rm bi}}q_{Hi}^2\right)g_3^2\right] \nonumber \\
\frac{d \chi}{dt}&=\frac{\chi}{16\pi^2}\left[\left(\frac{41}{6}+\frac{10}{9}N_{\rm bi}\right)g_Y'^2+\left(\frac{20}{3}\sum_{i=1}^{N_{\rm bi}}q_{Hi}^2\right)g_H^2\right] \nonumber \\
&+\frac{1}{(16\pi^2)^2}\left[\left(\frac{199}{18}+\frac{35}{54}N_{\rm bi}\right)\frac{g_Y'^4\chi}{1-\chi^2}+\left(\frac{10}{9}\sum_{i=1}^{N_{\rm bi}}q_{Hi}\right)\frac{1+\chi^2}{1-\chi^2}g_Y'^3g_H+\left(20\sum_{i=1}^{N_{\rm bi}}q_{Hi}^2\right)\frac{g_Y'^2g_H^2\chi}{1-\chi^2}\right. \nonumber \\
&\left.+\left(\frac{9}{2}+\frac{3}{2}N_{\rm bi}\right)\chi g_Y'^2g_2^2+\left(\frac{44}{3}+\frac{16}{9}N_{\rm bi}\right)\chi g_Y'^2g_3^2+\left(6\sum_{i=1}^{N_{\rm bi}}q_{Hi}\right)g_Y'g_Hg_2^2+\left(\frac{-32}{3}\sum_{i=1}^{N_{\rm bi}}q_{Hi}\right)g_Y'g_Hg_3^2\right. \nonumber\\
&\left.+\left(20\sum_{i=1}^{N_{\rm bi}}q_{Hi}^4\right)\frac{g_H^4\chi}{1-\chi^2}+\left(6\sum_{i=1}^{N_{\rm bi}}q_{Hi}^2\right)g_H^2g_2^2\chi+\left(16\sum_{i=1}^{N_{\rm bi}}q_{Hi}^2\right)g_H^2g_3^2\chi-\frac{17}{6}g_Y'^2y_t^2\chi \right] \nonumber \\
\frac{d g_2}{dt}&=\frac{g_2^3}{16\pi^2}\left(-\frac{19}{6}+\frac{2}{3}N_{\rm bi}\right) \nonumber \\
&+\frac{g_2^3}{(16\pi^2)^2}\left[\left(\frac{3}{2}+\frac{1}{2}N_{\rm bi}\right)\frac{g_Y'^2}{1-\chi^2}+\left(2\sum_{i=1}^{N_{\rm bi}}q_{Hi}^2\right)\frac{g_H^2}{1-\chi^2}+\left(2\sum_{i=1}^{N_{\rm bi}}q_{Hi}\right)\frac{g_Y'g_H\chi}{1-\chi^2}\right. \nonumber \\
&\left.+\left(\frac{35}{6}+\frac{49}{6}N_{\rm bi}\right)g_2^2+\left(12\right)g_3^2-\frac{3}{2}y_t^2\right] \nonumber \\
\frac{d g_3}{dt}&=\frac{g_3^3}{16\pi^2}\left(-7+\frac{2}{3}N_{\rm bi}\right) \nonumber \\
&+\frac{g_3^3}{(16\pi^2)^2}\left[\left(\frac{11}{6}+\frac{2}{9}N_{\rm bi}\right)\frac{g_Y'^2}{1-\chi^2}+\left(2\sum_{i=1}^{N_{\rm bi}}q_{Hi}^2\right)\frac{g_H^2}{1-\chi^2}+\left(-\frac{4}{3}\sum_{i=1}^{N_{\rm bi}}q_{Hi}\right)\frac{g_Y'g_H\chi}{1-\chi^2}\right. \nonumber \\
&\left.+\left(\frac{9}{2}\right)g_2^2+\left(-26+\frac{38}{3}N_{\rm bi}\right)g_3^2-2y_t^2\right] \nonumber \\
\frac{d y_t}{dt}&=\frac{y_t}{16 \pi ^2} \left(-\frac{17}{12}\frac{g_Y'^2}{1-\chi2}-\frac{9}{4}g_2^2-8g_3^2+\frac{9}{2}y_t^2\right) \nonumber \\
\frac{d \lambda}{dt}& =\frac{1}{16\pi^2}\left(24 \lambda^{2} -3\lambda \frac{g_Y'^{2}}{1-\chi^2} -9\lambda g_{2}^{2}  +12 \lambda y_t^2+\frac{3}{8}  \frac{g_Y'^{4}}{(1-\chi^2)^2} +\frac{3}{4}  \frac{g_Y'^{2}}{1-\chi^2} g_{2}^{2} +\frac{9}{8} g_{2}^{4} -6y_t^4\right)\nonumber 
\end{align}


\begin{thebibliography}{99}

\bibitem{susygut1}
P. Langacker, ``Precision Tests Of The Standard Model,'' in Proceedings of the PASCOS90 Symposium, (World Scientific, 1990).

\bibitem{Ellis:1990wk} 
  J.~R.~Ellis, S.~Kelley and D.~V.~Nanopoulos,
  Phys.\ Lett.\ B {\bf 260}, 131 (1991).
 
 \bibitem{Amaldi:1991cn} 
  U.~Amaldi, W.~de Boer and H.~Furstenau,
  Phys.\ Lett.\ B {\bf 260}, 447 (1991).

\bibitem{Langacker:1991an} 
  P.~Langacker and M.~x.~Luo,
  Phys.\ Rev.\ D {\bf 44}, 817 (1991).

\bibitem{Giunti:1991ta} 
  C.~Giunti, C.~W.~Kim and U.~W.~Lee,
  Mod.\ Phys.\ Lett.\ A {\bf 6}, 1745 (1991).






\bibitem{Murayama:1991ah} 
  H.~Murayama and T.~Yanagida,
  Mod.\ Phys.\ Lett.\ A {\bf 7}, 147 (1992).


\bibitem{Bachas:1995yt} 
  C.~Bachas, C.~Fabre and T.~Yanagida,
  Phys.\ Lett.\ B {\bf 370}, 49 (1996)
  [hep-th/9510094].
  
  
  

\bibitem{Redondo:2008zf} 
  J.~Redondo,
  arXiv:0805.3112 [hep-ph].
  
  
  
  
  
  
  
  
  
  
\bibitem{Takahashi:2016iph} 
  F.~Takahashi, M.~Yamada and N.~Yokozaki,
  Phys.\ Lett.\ B {\bf 760}, 486 (2016)
  [arXiv:1604.07145 [hep-ph]].
  
  
  
  
  
  
  
  
  
  
  
  \bibitem{Marinelli:1982dg} 
  M.~Marinelli and G.~Morpurgo,
  Phys.\ Rept.\  {\bf 85}, 161 (1982).
  
  \bibitem{Davidson:1993sj} 
  S.~Davidson and M.~E.~Peskin,
  Phys.\ Rev.\ D {\bf 49}, 2114 (1994)
  [hep-ph/9310288].

\bibitem{Davidson:2000hf} 
  S.~Davidson, S.~Hannestad and G.~Raffelt,
  JHEP {\bf 0005}, 003 (2000)
  [hep-ph/0001179].


  
  \bibitem{Lee:2002sa} 
  I.~T.~Lee {\it et al.},
  Phys.\ Rev.\ D {\bf 66}, 012002 (2002)
  [hep-ex/0204003].
  
  
  
\bibitem{Dubovsky:2003yn} 
  S.~L.~Dubovsky, D.~S.~Gorbunov and G.~I.~Rubtsov,
  JETP Lett.\  {\bf 79}, 1 (2004)
  [Pisma Zh.\ Eksp.\ Teor.\ Fiz.\  {\bf 79}, 3 (2004)]
  [hep-ph/0311189].


  \bibitem{Kim:2007zzs} 
  P.~C.~Kim, E.~R.~Lee, I.~T.~Lee, M.~L.~Perl, V.~Halyo and D.~Loomba,
  Phys.\ Rev.\ Lett.\  {\bf 99}, 161804 (2007).


  \bibitem{McDermott:2010pa} 
  S.~D.~McDermott, H.~B.~Yu and K.~M.~Zurek,
  Phys.\ Rev.\ D {\bf 83}, 063509 (2011)
  [arXiv:1011.2907 [hep-ph]].
 
 
 \bibitem{Essig:2013lka} 
  R.~Essig {\it et al.},
  ``Working Group Report: New Light Weakly Coupled Particles,''
  arXiv:1311.0029 [hep-ph].

 
  
\bibitem{Dolgov:2013una} 
  A.~D.~Dolgov, S.~L.~Dubovsky, G.~I.~Rubtsov and I.~I.~Tkachev,
  Phys.\ Rev.\ D {\bf 88}, no. 11, 117701 (2013)
  [arXiv:1310.2376 [hep-ph]].

  
\bibitem{Vogel:2013raa} 
  H.~Vogel and J.~Redondo,
  JCAP {\bf 1402}, 029 (2014)
  [arXiv:1311.2600 [hep-ph]].


\bibitem{Moore:2014yba} 
  D.~C.~Moore, A.~D.~Rider and G.~Gratta,
  Phys.\ Rev.\ Lett.\  {\bf 113}, no. 25, 251801 (2014)
  [arXiv:1408.4396 [hep-ex]].

  
\bibitem{Kamada:2016qjo} 
  A.~Kamada, K.~Kohri, T.~Takahashi and N.~Yoshida,
  arXiv:1604.07926 [astro-ph.CO].

  
  
  
  
  
  
  
  
  


  
\bibitem{Cline:2012is} 
  J.~M.~Cline, Z.~Liu and W.~Xue,
  Phys.\ Rev.\ D {\bf 85}, 101302 (2012)
  [arXiv:1201.4858 [hep-ph]].
  

\bibitem{Foot:2014uba} 
  R.~Foot and S.~Vagnozzi,
  Phys.\ Rev.\ D {\bf 91}, 023512 (2015)
  [arXiv:1409.7174 [hep-ph]].
  
\bibitem{Foot:2014osa} 
  R.~Foot and S.~Vagnozzi,
  Phys.\ Lett.\ B {\bf 748}, 61 (2015)
  [arXiv:1412.0762 [hep-ph]].

\bibitem{Foot:2016wvj} 
  R.~Foot and S.~Vagnozzi,
  JCAP {\bf 1607}, no. 07, 013 (2016)
  [arXiv:1602.02467 [astro-ph.CO]].

  
\bibitem{Cardoso:2016olt} 
  V.~Cardoso, C.~F.~B.~Macedo, P.~Pani and V.~Ferrari,
  JCAP {\bf 1605}, no. 05, 054 (2016)
  [arXiv:1604.07845 [hep-ph]].
  
\bibitem{Foot:2014mia} 
  R.~Foot,
  Int.\ J.\ Mod.\ Phys.\ A {\bf 29}, 1430013 (2014)
  [arXiv:1401.3965 [astro-ph.CO]].


  
\bibitem{Holdom:1985ag}
  B.~Holdom,
  Phys.\ Lett.\  {\bf 166B}, 196 (1986).
  
  
\bibitem{Glashow:1985ud}
  S.~L.~Glashow,
  Phys.\ Lett.\  {\bf 167B}, 35 (1986).


\bibitem{Carlson:1987si}
  E.~D.~Carlson and S.~L.~Glashow,
  Phys.\ Lett.\ B {\bf 193}, 168 (1987).
  
  
  
  
  
  
  
  
  
  
  
  
  \bibitem{Babu:1996vt} 
  K.~S.~Babu, C.~F.~Kolda and J.~March-Russell,
  Phys.\ Rev.\ D {\bf 54}, 4635 (1996)
  [hep-ph/9603212].

\bibitem{Lyonnet:2013dna} 
  F.~Lyonnet, I.~Schienbein, F.~Staub and A.~Wingerter,
  Comput.\ Phys.\ Commun.\  {\bf 185}, 1130 (2014)
  [arXiv:1309.7030 [hep-ph]].

\bibitem{Lyonnet:2016xiz} 
  F.~Lyonnet and I.~Schienbein,
  arXiv:1608.07274 [hep-ph].











\bibitem{DelNobile:2015bqo} 
  E.~Del Nobile, M.~Nardecchia and P.~Panci,
  JCAP {\bf 1604}, no. 04, 048 (2016)
  [arXiv:1512.05353 [hep-ph]].



\bibitem{Chuzhoy:2008zy}
  L.~Chuzhoy and E.~W.~Kolb,
  JCAP {\bf 0907} (2009) 014
  [arXiv:0809.0436 [astro-ph]].







\bibitem{Davoudiasl:2004be} 
  H.~Davoudiasl, R.~Kitano, T.~Li and H.~Murayama,
  Phys.\ Lett.\ B {\bf 609}, 117 (2005)
  [hep-ph/0405097].


\bibitem{Patt:2006fw} 
  B.~Patt and F.~Wilczek,
  hep-ph/0605188.

\bibitem{Kim:2006af} 
  Y.~G.~Kim and K.~Y.~Lee,
  Phys.\ Rev.\ D {\bf 75}, 115012 (2007)
  [hep-ph/0611069].
  
  \bibitem{Barger:2007im} 
  V.~Barger, P.~Langacker, M.~McCaskey, M.~J.~Ramsey-Musolf and G.~Shaughnessy,
  Phys.\ Rev.\ D {\bf 77}, 035005 (2008)
   [arXiv:0706.4311 [hep-ph]].

\bibitem{Kim:2008pp} 
  Y.~G.~Kim, K.~Y.~Lee and S.~Shin,
  JHEP {\bf 0805}, 100 (2008)
   [arXiv:0803.2932 [hep-ph]].
  


\bibitem{Beniwal:2015sdl} 
  A.~Beniwal, F.~Rajec, C.~Savage, P.~Scott, C.~Weniger, M.~White and A.~G.~Williams,
  Phys.\ Rev.\ D {\bf 93}, no. 11, 115016 (2016)
  [arXiv:1512.06458 [hep-ph]].


\bibitem{Akerib:2016vxi} 
  D.~S.~Akerib {\it et al.},
  arXiv:1608.07648 [astro-ph.CO].



\bibitem{Nomura:2008ru} 
Y.~Nomura and J.~Thaler,
  Phys.\ Rev.\ D {\bf 79}, 075008 (2009)
  [arXiv:0810.5397 [hep-ph]].

\bibitem{Peccei:1977hh} 
  R.~D.~Peccei and H.~R.~Quinn,
  Phys.\ Rev.\ Lett.\  {\bf 38}, 1440 (1977).

\bibitem{Weinberg:1977ma} 
  S.~Weinberg,
  Phys.\ Rev.\ Lett.\  {\bf 40}, 223 (1978).

\bibitem{Wilczek:1977pj} 
  F.~Wilczek,
  Phys.\ Rev.\ Lett.\  {\bf 40}, 279 (1978).

\bibitem{Kim:1979if} 
  J.~E.~Kim,
  Phys.\ Rev.\ Lett.\  {\bf 43}, 103 (1979).
  
\bibitem{Shifman:1979if} 
  M.~A.~Shifman, A.~I.~Vainshtein and V.~I.~Zakharov,
  Nucl.\ Phys.\ B {\bf 166}, 493 (1980).

\bibitem{wgc}
  N.~Arkani-Hamed, L.~Motl, A.~Nicolis and C.~Vafa,
  JHEP {\bf 0706}, 060 (2007)
  [hep-th/0601001].
  
  \bibitem{Luo:2002iq} 
  M.~x.~Luo and Y.~Xiao,
  Phys.\ Lett.\ B {\bf 555}, 279 (2003)
  [hep-ph/0212152].



\end{thebibliography}
\end{document}